# On The Evolution of PON-Based FTTH Solutions

Kyeong Soo Kim
Advanced System Technology, STMicroelectronics
Stanford Networking Research Center, Packard Building, Room 073, Stanford, CA 94305 (kks@stanford.edu).

*Abstract*—Passive Optical Network (PON)-based Fiber-To-The-Home (FTTH) are promising solutions that can break through the economic barrier of traditional point-to-point solutions. Once fibers are deployed with PON-based FTTH solutions, it becomes critical how to migrate to Wavelength Division Multiplexing (WDM)–PON because Time Division Multiplexing (TDM) used in current PON solutions cannot exploit the huge bandwidth of the optical fibers and therefore will not be able to meet ever-increasing demands for higher bandwidth by future network applications. In this paper we review and compare the current PON-based FTTH solutions, ATM-PON (APON) and Ethernet PON (EPON), and provide a possible evolution scenario to future WDM-PON.

## I. Introduction

FIBER-TO-THE-HOME (FTTH) has been considered an ideal solution for access networks since the invention of optical fiber communications because of huge capacity, small size and lightness, and immunity to electromagnetic interference of optical fibers. For example, it is shown that the information capacity of an optical fiber can exceed 100 Tb/s under propagation nonlinearity for a typical Dense Wavelength Division Multiplexing (DWDM) system with coherent detection [1]. Because optical fibers are widely used in backbone networks, Wide Area Networks (WANs), and Metropolitan Area Networks (MANs), and are also being deployed in Local Area Networks (LANs) with the introduction of new optical Ethernet standards, the implementation of the FTTH in access networks, which are also called 'the last mile', will complete all-optical-network revolution.

Back in early 1990s, International Telegraph and Telephone Consultative Committee (CCITT)[1] issued several I series recommendations for Broadband aspects of Integrated Services Digital Network (B-ISDN) based on Asynchronous Transfer Mode (ATM) technology [2]. Early B-ISDN standards specified two User-Network Interfaces (UNIs) at speeds of 155.520 Mbps and 622.020 Mbps, respectively, which were the first FTTH solution ever standardized. The original B-ISDN UNIs, however, have never been successfully deployed due to their high cost and lack of killer applications then for those bandwidths.

To address these issues and expedite the introduction of FTTH, Passive Optical Network (PON)-based solutions, ATM-PON (APON) and Ethernet PON (EPON), have been recently proposed. These solutions are based on the common network architecture, *i.e.*, PON, but adopt different transfer technologies to support integrated services and multiple protocols. Because these solutions, especially APON, are considered as upgrades of the Asynchronous Digital Subscriber Lines (ADSLs) and therefore strongly supported by Incumbent Local Exchange Carriers (ILECs), they will serve more residential end-users than any other types of FTTH solutions.

Once fibers are deployed in the field, it would be a critical issue how to migrate to WDM-PON because current Time Division Multiplexing (TDM)-based PON solutions cannot exploit the huge bandwidth of the optical fibers and therefore will have difficulties in meeting ever-increasing demands for higher bandwidth by future network applications.

In this paper we focus our discussions of FTTH on the PON-based solutions. First, in Section II we review two current Time Division Multiplexing (TDM)-based PON solutions, APON and EPON, and summarize their advantages and disadvantages in various aspects. Then in Section III we discuss a possible evolution scenario from APON/EPON to WDM-PON with issues to be addressed at each step of the scenario. Section IV concludes this paper.

## II. Overview of Current PON-Based FTTH Solutions

### A. APON

Around 1995 several telecommunications operators formed the Full Services Access Network (FSAN) initiative [3] to break through the economic barrier by large-scale introduction of broadband access networks through the definition of basic set of common requirements, which now become ITU-T Recommendations G.983 series.

Two key technologies adopted by FSAN in achieving this goal are ATM and PON. As is well known, the ATM is a switching and transfer methodology that can handle different types of services like voice, video, and data in the same network with guaranteed Quality of Service (QoS). In traditional ATM-based access networks statistical multiplexers, also called edge/access switches, are used to multiplex many incoming streams from end-users into a single outgoing stream to networks. Because the statistical multiplexers are active systems that need power and protections and to be deployed between customer premises (houses in case of FTTH) and a Central Office (CO), maintaining them, which is usually done by craftsmen,

---

[1] CCITT was replaced by International Telecommunication Union Telecommunication standardization sector (ITU-T) on March 1, 1993.

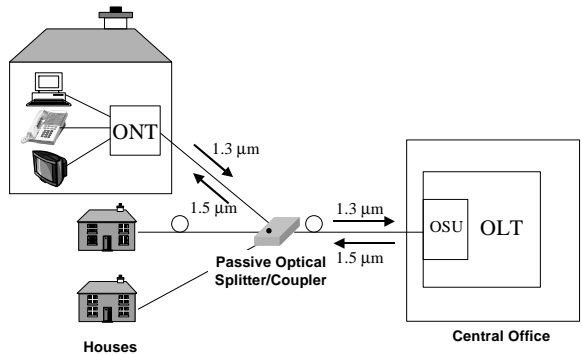

Fig. 1. PON-based FTTH system architecture.

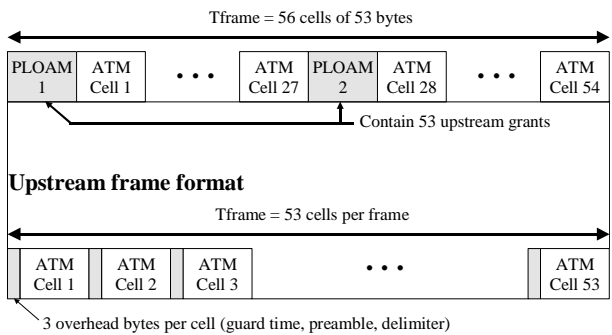

Fig. 2. APON frame formats.

contributes to the high cost of fiber-based access networks.

The PON solves this problem by replacing the active systems outside the plant with passive optical splitters/couplers. Fig. 1 shows PON-based FTTH system architecture where a PON connects several Optical Network Terminations (ONTs) in customer premises and an Optical Subscriber Unit (OSU) at Optical Line Termination (OLT) in CO. Coarse WDM (CWDM) is used to provide full-duplex bidirectional communications with 1300 nm band for upstream and 1500 nm band for downstream transmission.

Because ATM cells from ONTs are combined into one cell stream to the OSU after the optical splitter/coupler, there may be collision among upstream cells from different ONTs unless there is a Media Access Control (MAC) protocol to coordinate their transmission. So TDM-based MAC is used with a frame structure for collision-free upstream transmission on the PON. Note that in downstream transmission there is no such a conflict.

Fig. 2 shows APON frame formats. The upstream channel is divided into 53 slots of 56 bytes at 155.520 Mbps, while the downstream cell stream is divided into frames of 56 cells at 155.520 Mbps.[2]

At the beginning and the middle of a downstream frame Physical Layer Operation and Maintenance (PLOAM) cells are inserted, which are used by the OSU to control and communicate with the ONTs. Each PLOAM cell contains 27 (26 for the second PLOAM cell) grant fields and a 12-byte message field. The messages may be either broadcast, idle, or for a specific ONT. Grants are used to control the upstream data transmission, while messages are used to control the operation of the ONTs. There are basically 3 types of grants, DATA, UNASSIGNED, and PLOAM grants.[3]

In each upstream slot a 3-byte overhead header and a 53-byte ATM cell may be transmitted by one ONT. Which ONT is allowed to transmit a cell in that time slot is decided by the value of the grant field in the downstream PLOAM cell for that particular time slot. There is a one-to-one correspondence between the grant fields in the first and second PLOAM cells of a downstream frame and slots in an upstream frame. During the process of ranging, which will be described shortly, each ONT is assigned unique GRANT_ID (for either user data or PLOAM cells). Whenever the ONT sees a matching GRANT_ID in a grant field for a certain time slot, it transmits at that time slot a cell. Depending on the grant type, the transmitted cell is a PLOAM cell (if it is a PLOAM GRANT), a user ATM cell (if it is a DATA GRANT and the ONT has data to transmit) or an idle cell (if it is a DATA GRANT but the ONT has no data to transmit). If the GRANT_ID does not match that of the ONT or an UNASSIGNED GRANT_ID is received, the ONT does not transmit a cell. With this mechanism the OSU can not only prevent collision among upstream cells but also assign upstream bandwidth to each ONT by allocating proper number of grants per frame to it.

In the above process all ONTs must agree on the time slot boundaries. Otherwise, due to the differences in the fiber lengths connecting each ONT and the splitter, the upstream cells arriving at the OSU will overlap and the corresponding cells will be lost. Ranging ensures that cells do not overlap. During the ranging the physical distance between each ONT and the OLT is measured. After the ranging, the OLT puts equalization delay at each ONT, and by doing so all ONTs are logically placed at the same distance from the OLT irrespective of their physical distances.

Another key technology is burst mode reception at the OSU. Optical receivers set their decision threshold level at the optimum point to decide whether an input signal is 0 or 1. For conventional point-to-point communications, such as Synchronous Optical NETwork (SONET)/Synchronous Digital Hierarchy (SDH) systems, this level is fixed as long as the laser power and the length of the fiber is the same. In case of PON this optimum level will differ depending on which ONT transmitted the cell. This is because as the propagation length and transmitter power are different for each ONT, the strength of the optical signal at the point of reception at the OSU will differ with each ONT. So for each burst of data from an individual ONT, the receiver's decision threshold level must be reset[4]. In addition to power level, transmission clocks of ONTs are different one another. Therefore reset and synchronization should be done within overhead period of each upstream slot, which is a very challenging task especially at high speed.

---

[2] ITU-T G.983.1 also specifies 622.020 Mbps downstream frame format.

[3] Other types exist in the ITU-T G.983.1 specifications.

TABLE I. APON VS. EPON

|  | APON | EPON |
|---|---|---|
| Standard Body | ITU-T/FSAN | IEEE |
| Speed | 155/622 Mbps | 1 Gbps |
| Protocol overhead For IP services | Large | Small |
| Scalability | Low | High (up to 10 Gbps) |
| Initial Capital Expenditure | Low | Mid to High |
| Service Integration | Good | Good |
| Strong Players | ILECs | CLECs |

*B. EPON*

EPON is one of the solutions considered by new IEEE 802.3ah Ethernet in the First Mile (EFM) Task Force [4], focusing on direct support of Ethernet services. EPON is currently under standardization[5], so we will not discuss details of its specifications here.

Since most of key technologies are borrowed from the APON, EPON is very similar to APON in its basic operations. Like APON, EPON uses CWDM and TDM to provide bi-directional, point-to-point communications over a fiber and maintains frame structure for both upstream and downstream. One significant distinction is that in EPON variable-length frames (IEEE 802.3 Ethernet format) are used. Since EPON uses Ethernet as its transfer protocol, it has the following benefits compared to the APON:

- Low protocol overhead
- Higher bandwidth
- Lower costs
- Broader service capabilities
- Easy integration of LANs into future optical-Ethernet-based WANs

Table 1 summarizes the comparison of APON and EPON in several key aspects.

## III. EVOLUTION TOWARDS WDM-PON

Main purpose of the current PON solutions is to deploy the fiber in the last mile as economically as possible. Because of relatively higher cost of WDM components, TDM has been adopted to provide point-to-point connections between OLT and ONT. As we reviewed in Section II, however, the current TDM-based MAC protocol is highly complicated and difficult to upgrade for higher bandwidth. As the cost of WDM components rapidly drops and bandwidth-intensive network applications continue to emerge, in the near future upgrade of current TDM-PON to WDM-PON will be unavoidable.

---
[4] Another possible solution is to adjust the laser transmit power at the ONTs so that the power levels detected at the OSU is the same for all ONTs. But this solution has not been adopted by ITU-T G.983.1.
[5] The goal is to have baseline documents by March 2002.

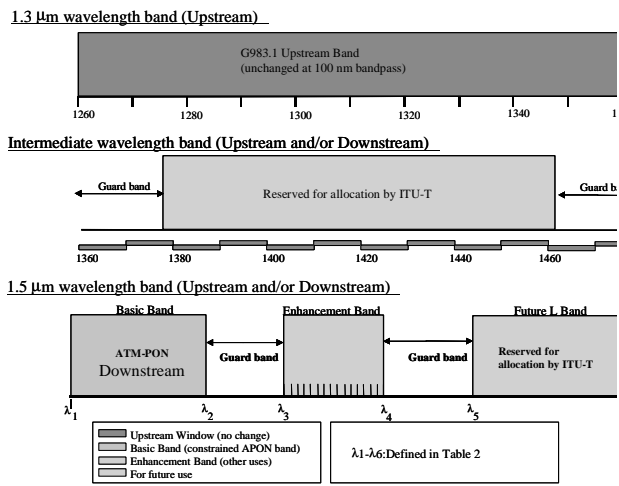

Fig. 3. Wavelength allocation by ITU-T G.983.3 [5].

One important issue is an evolution scenario to guarantee coexistence of and interoperation between new and old PON systems at each step. By separating bands for the original PON (basic band) and WDM-PON (enhancement band) as shown in Fig. 3, ITU-T G.983.3 makes clear that the introduction of new services by WDM techniques shouldn't disturb the basic TDM-based APON systems [5]. In the following we consider one example scenario for this evolution consisting of three steps.

*A. Broadcast Video Overlay by WDM*

At an earlier stage, broadcast video overlay would be a major driving force for WDM-PON because while cable TV service providers provide data services via cable modems in addition to broadcast video services, telecommunication operators cannot deliver comparable video broadcasting with current TDM-based PON systems.

Fig. 4 shows a system configuration to deliver broadcast video signal using enhancement band with the original basic band signal on the same PON by WDM, where CWDM1 separates the 1300 nm and 1500 nm bands, and CWDM2 splits the basic and enhancement bands within the 1500 nm range. As an alternative FDM video overlay technique may be used as shown in [5].

*B. Coexistence of TDM-PON and WDM-PON Systems*

If TDM-based systems cannot meet some users' demands for bandwidth due to new type of digital services like high-quality streaming video and audio, virtual reality, and new graphic-intensive web interfaces, it is needed to upgrade part of the systems to WDM-enabled ones where DWDM technique is used to provide a logical point-to-point link between OLT and ONT instead of TDM MAC protocol.

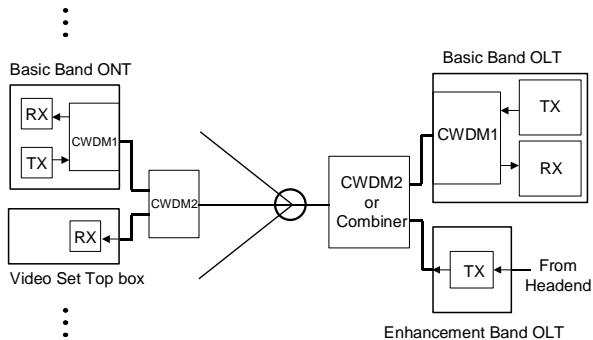

Fig. 4. Broadcast video overly by WDM.

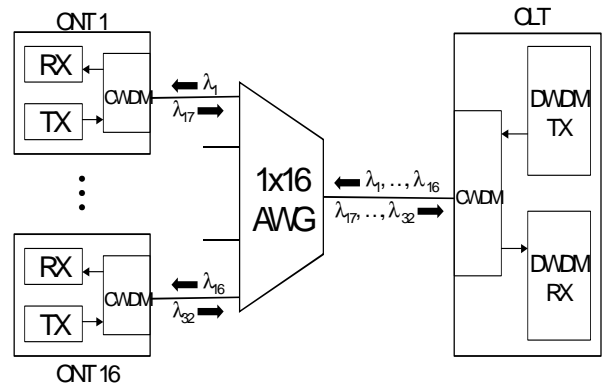

Fig. 6. Full WDM-PON with AWG.

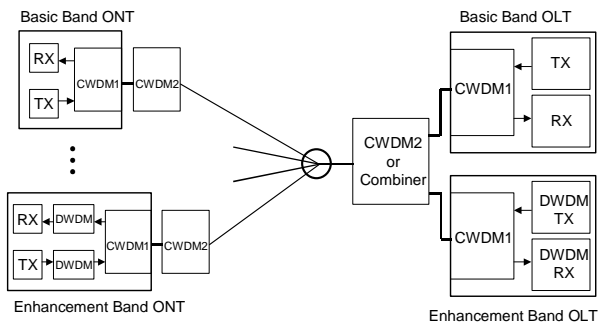

Fig. 5. Coexistence of TDM-PON and WDM-PON systems.

In this case new WDM-enabled ONTs and OLTs will share the same PON infrastructure with the existing TDM-PON systems. These first-generation WDM-PON ONTs will be equipped with CWDM filters (CWDM2) to separate enhancement band (for downstream) and intermediate wavelength band (for upstream, if needed) as well as DWDM filters as shown in Fig. 5.

*C. Full WDM-PON with AWGs*

If there is no need to support TDM-PON systems any longer, it would be better to replace power splitters/couplers with Arrayed Waveguide Gratings (AWGs) or Waveguide Grating Routers (WGRs), which eliminate the need for having a DWDM filter at each ONT as shown in Fig. 5. The system configuration for this case is shown in Fig. 6, and this will be the last step in the evolution towards WDM-PON.

Although it is conceptually simple, the use of AWG outside the plant has many problems of its own. For example, since AWGs are very sensitive to temperature changes, this puts severe constraints on channel spacing. Otherwise it would be necessary to monitor the passband wavelengths of the AWG and to tune the DWDM sources [7].

## IV. CONCLUDING REMARKS

In this paper we have provided an overview of current PON-based FTTH solutions, APON and EPON, and an example scenario for evolution towards WDM-PON.

In the literature there have been many researches focusing on WDN-PON, but few of them deal with migration strategies that enables smooth and economical transition to WDM-PON. The steps discussed in Section III are given as examples, and more or less steps may be needed in actual situations. For example, one may need another step between Fig. 5 and Fig. 6, where TDM-PON and WDM-PON systems coexist on the PON with both power splitters and AWGs.

In addition to the evolution scenarios and the corresponding system architectures, issues like the use of tunable lasers at ONTs, AWG wavelength tracking, and protocols for DWDM link need further study.

ACKNOWLEDGMENT

The author thanks AST Optical Networking R&D Team for useful discussions and their encouragement.